\documentstyle[prl,aps,twocolumn,epsf]{revtex}
\def\break#1{\pagebreak \vspace*{#1}}
\begin{document}
\draft
\title{ Voltage-biased quantum wire with impurities }
\author{Reinhold Egger and Hermann Grabert}
\address{
Fakult\"at f\"ur Physik, Albert-Ludwigs-Universit\"at, 
Hermann-Herder-Stra{\ss}e 3,
D-79104 Freiburg, Germany}
\maketitle
\widetext
\begin{abstract}
The bosonization technique to describe correlated electrons in
a one-dimensional quantum wire containing impurities is extended to include 
an applied voltage source. The external reservoirs are shown to
lead to a boundary condition for the boson phase fields.
We use the formalism to investigate the channel conductance,
electroneutrality, and charging effects.
\end{abstract}


\narrowtext

The puzzling physical properties of one-dimensional 
correlated fermions at low temperatures can  most
conveniently be described within the bosonization technique.
This method \cite{luther,emery,haldane}
allows for an exact treatment of Coulomb 
interactions that hamper most other theoretical 
approaches. In this Letter, we describe how a
voltage bias can properly be incorporated in terms of a
boundary condition for the bosonized phase fields.
The formalism is similar in spirit to Landauer's
approach developed for noninteracting
electrons \cite{landauer}.  The capacity of our
concept is demonstrated for a single impurity embedded
into the correlated one-dimensional (1D) electron liquid,
a problem that has attracted considerable 
theoretical interest \cite{kane92,glaz}, and is beginning to 
find experimental realizations \cite{jap}.

In order to describe Coulomb interactions adequately, one has to specify the
setup under consideration. If one deals with a 1D channel in 
heterostructures, a ``quantum wire''\cite{gog}, the interactions
are usually screened due to the presence of metallic gates near the
channel. This leads to a Luttinger liquid\cite{haldane}
characterized by an interaction constant $g$ (we only discuss the
spinless single-channel case in the following). The noninteracting  case
corresponds to $g=1$, and the presence of (repulsive) Coulomb interactions
implies $g<1$. Quantities of principal interest are the channel 
conductance in the presence of interactions and
the capacitance $C=Q/U$ of a junction or impurity 
($U$ is the two-terminal voltage, 
and $Q$ denotes the charge on the junction).
We present an approach that allows one to address these problems for
finite voltage at 
arbitrary interaction strength and junction transmittance.

Our treatment is based on the standard bosonization 
approach\cite{luther,emery,haldane},
which is applicable in the low-energy regime where only excitations
near the Fermi surface are relevant. The electron creation operator 
can be expressed in terms of boson phase fields $\theta(x)$ and
$\phi(x)$,
\[
\psi^\dagger(x) = \sqrt{\frac{\omega_c}{2\pi v_F}} \,\sum_{p=\pm}
\exp\left[ i p k_F x +  i\sqrt{\pi} [ p\theta(x) + \phi(x)]\right ]\;,
\]
where $\omega_c=v_F k_F$ is the proper bandwidth cutoff
for the linearized dispersion relation employed in the 
bosonization (we put 
$\hbar =1$).
The phase fields obey the equal-time 
\break{1.0in}
commutation relations
\[
 [ \phi(x) , \theta(x')] =  -(i/2) \, {\rm sgn} (x-x') \;,
\]
such that the canonical momentum for the $\theta$ field is 
$\Pi=\partial_x \phi$. The boson representation for
the  electron density operator  is then given by
\begin{equation} \label{bosondens}
\rho(x) = \frac{k_F}{\pi} + \frac{1}{\sqrt{\pi}} \partial_x \theta(x)
+ \frac{k_F}{\pi} \cos[2k_F x+ 2\sqrt{\pi} \theta(x)] \;.
\end{equation}
The first term is the background charge, the second term stands
for the sum of right- and left-moving densities $\rho_{\pm}$, and the
last term describes interference between right- and 
left-movers \cite{emery}.
The clean Luttinger liquid is
described by the Euclidean action\cite{haldane}
\begin{equation}\label{lutt}
S_0 =\frac{v_F}{2} \int dx d\tau\left [\frac{1}{v_F^2} 
(\partial_\tau \theta)^2 + 
\frac{1}{g^2} (\partial_x \theta)^2\right ]\;,
\end{equation}
where $v_F$ is the Fermi velocity and $g\leq 1$ the interaction constant.
A short-ranged impurity at $x=0$ results in  the generic 
contribution\cite{kane92}
\begin{equation} \label{impurity}
S_I = V \int d\tau \cos[2\sqrt{\pi}\theta(0,\tau)]\;,
\end{equation}
where the dimensionless impurity strength $\lambda=\pi V/\omega_c$
tunes the junction resistance.

Now let us consider a 1D quantum wire coupled
to external reservoirs, see Fig.~1. The coupling of 
the 1D channel to the 2D or 3D reservoirs is assumed 
to occur by adiabatic widening of the channel.
This models ideal reservoirs in the sense of Landauer 
\cite{landauer}.
The reservoirs are macroscopic, totally incoherent  
devices kept at fixed chemical potential, and there are 
no reflections of particles entering the reservoirs. 
The left (right) reservoir at $x\to -\infty$ ($x\to \infty$)
has chemical potential $\mu_{-\infty}$ ($\mu_\infty$), and
the difference in chemical potentials is the two-terminal voltage
$U = (\mu_{-\infty} - \mu_\infty)/e$. 
The mean chemical potential $(\mu_{-\infty}+\mu_\infty)/2$ 
corresponds to the background charge $k_F/\pi$. 
\vbox{
\epsfysize=4.2cm
\epsffile{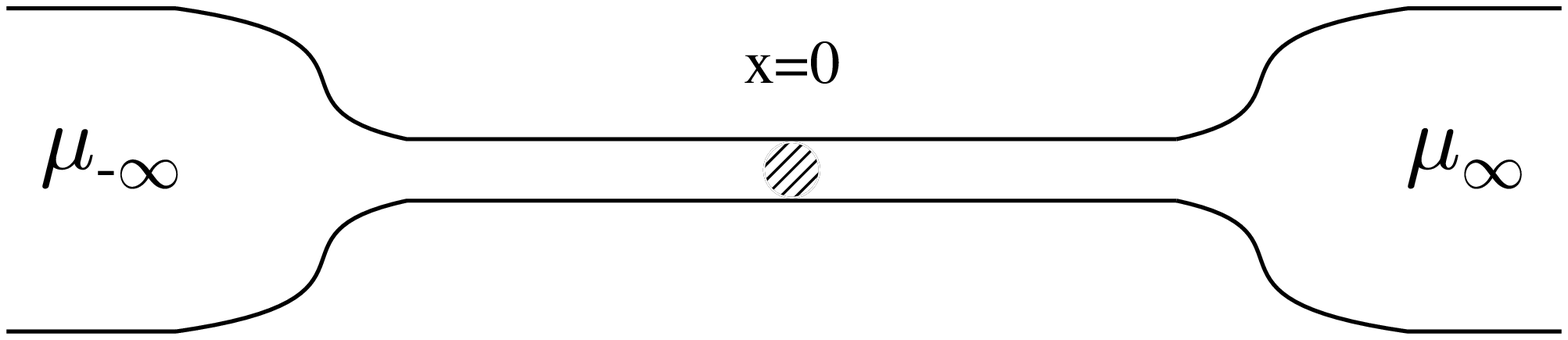}
\begin{figure}
\caption[]{\label{fig1}
General setup studied in this work:
A 1D quantum wire is coupled to 3D reservoirs, which are held
at chemical potentials $\mu_{-\infty}$
and $\mu_\infty$. The left reservoir injects right-movers at the
stationary excess density $\langle \rho_+ \rangle$, and the right
one injects left-movers with density $\langle \rho_-\rangle$.
The striped area stands for a scatterer at $x=0$.}
\end{figure}}
Applying a voltage means that one has a non-equilibrium influx of 
currents from the reservoirs: The reservoir at $x\to -\infty$ 
injects right-movers at some stationary
excess density, and similarly the density of left-movers at
$x\to \infty$ is diminished,
\begin{equation}\label{bc1}
\langle\rho_\pm (x)\rangle  = \pm \frac{eU}{4\pi v_F} \qquad {\rm as}\;
x\to \mp \infty\;.
\end{equation} 
While in a clean system the densities of right- and left-movers
imposed by the reservoirs spread homogeneously along the wire,
in the presence of scatterers one does not know beforehand what 
the densities of right- (left-) movers at $x\to \infty$
($x\to -\infty$) are. These densities follow from our theory. 

The  boundary condition (\ref{bc1}), formulated in terms of the average 
densities of right- or left-moving fermions in the quantum wire
as the reservoirs are approached, can equivalently be expressed in terms of 
the $\theta$ field by noting that
\[
\rho_+ + \rho_-  = \frac{1}{\sqrt{\pi}} \partial_x \theta\;,\qquad
 \rho_+ - \rho_- = \frac{i}{v_F\sqrt{\pi}}  \partial_\tau \theta\;.
\]
The boundary condition for the boson phase field is therefore
\begin{equation} \label{bc}
\left ( \pm \frac{\partial}{\partial x} + \frac{i}{v_F}
 \frac{\partial}{\partial \tau} \right ) 
\langle\theta (x\to \mp \infty, \tau=0) \rangle
= \frac{eU}{2\sqrt{\pi} v_F} \;,
\end{equation}
where the time $\tau=0$ has been picked by convention.

To study the inhomogeneous quantum wire
 in presence of an external voltage, we consider the generating 
functional 
$Z(y,\mu) = \langle \exp [2\sqrt{\pi}\,i \mu \theta(y)] \rangle$.
We formally solve for $Z$ by introducing an auxiliary field 
$q(\tau)= 2\sqrt{\pi}\, \theta(x=0,\tau) $, with the
constraint enforced by a Lagrange multiplier field $\Lambda(\tau)$.
Then one has the effective action
\begin{eqnarray} 
 \nonumber
S_e[\theta,\Lambda,q]
 &=& \frac{v_F}{2} \int dx d\tau \left [ \frac{1}{v_F^2}(\partial_\tau
\theta )^2 + \frac{1}{g^2} (\partial_x \theta)^2 \right] \\
\label{seff}
&+&  V \int d\tau \, \cos q(\tau) -2\sqrt{\pi}\, i \mu\theta(y,0)\\
&+& i \int d\tau\, \Lambda(\tau)\,  [2\sqrt{\pi}
\,\theta(0,\tau) - q(\tau)]  \;. \nonumber
\end{eqnarray}
The $\theta$ part of this effective action is Gaussian and can therefore
be treated exactly by solving the classical Euler-Lagrange equations. 
One can always decompose the solution $\theta$ into
an homogeneous part $\theta_h$ for the equilibrium system ($U=0$),
and a particular solution
 $\theta_p$ subject to the boundary condition (\ref{bc}).  At the same time,
we decompose the field $\Lambda = \Lambda_h + \Lambda_p$, such that
\begin{equation}\label{parti}
 \frac{1}{v_F^2} \frac{\partial^2 \theta_p}{\partial \tau^2}  + 
\frac{1}{g^2} \frac{\partial^2 \theta_p}{\partial x^2} 
 = \frac{2\sqrt{\pi}\, i}{v_F} \, \delta(x) \Lambda_p(\tau)\;.
\end{equation}
The most general solution permitted by Eq.~(\ref{parti}) 
which fulfills Eq.~(\ref{bc})
requires a $\tau$ independent $\Lambda_p$ and takes the form
\begin{equation} \label{parsol}
\theta_p(x,\tau) =\frac{q_0}{2\sqrt{\pi}}
 - \frac{e\varphi}{2\sqrt{\pi}\,v_F} |x|
- i \tau \frac{e(U-\varphi)}{2\sqrt{\pi}} \;.
\end{equation}
The quantity $\varphi$ is related to the zero mode of the Lagrange 
multiplier field, $\Lambda_p=ie\varphi/2\pi g^2$.

With the boson propagators
\[
F(x,\omega) = \frac{\pi g}{|\omega|} \, \exp( -|g\omega x/v_F|) \;,
\]
the homogeneous part can be written in terms of the Fourier 
components $\Lambda_h(\omega)$, 
\[
\theta_h(x,\tau) = \frac{-i}{\sqrt{\pi}} \int
 \frac{d\omega}{2\pi}\, e^{i\omega \tau} 
\, [\Lambda_h(\omega) F(x,\omega) - \mu F(x-y,\omega) ].
\]
Inserting $\theta_h+\theta_p$ into Eq. (\ref{seff}), the action
becomes Gaussian in $\Lambda_h$, which can therefore easily be
integrated out. 
Adding the particular solution (\ref{parsol}) also onto
$q(\tau)$ in view of $q(\tau)=2\sqrt{\pi} \theta(0,\tau)$, 
one obtains for the generating functional
\begin{eqnarray}  \nonumber
&& \langle \exp [2\sqrt{\pi}i\mu\theta(x)] \rangle  =
W(x)^{\mu^2} \Bigl \langle e^{-i\mu e \varphi |x|/v_F} \\ && \;
\times 
\exp\left[i\mu \left( q_0 + \int \frac{d\omega}{2\pi}
q(\omega) \frac{F(x,\omega)}{F(0,\omega)}\right)\right] \Bigr\rangle\;,
\label{gen}
\end{eqnarray}
where the average over the zero modes $q_0$ and $\varphi$ of
the auxiliary fields and over the $q$ fluctuations $q(\tau)=(2\pi)^{-1}\int
d\omega q(\omega) \exp(i\omega\tau)$ has to be taken using the action
\begin{eqnarray} \nonumber
S&=&  \int \frac{d\omega}{2\pi} \frac{q(\omega) q(-\omega)}
{4 F(0,\omega)} 
+\frac{ e\varphi }{2\pi g^2} \int d\tau \, q(\tau)\\ \label{dissac}
&+& V \int d\tau \, \cos[q_0 - ie(U-\varphi)\tau +q(\tau) ] \;.
\end{eqnarray}
The function $W(x)=(1+|x|/\alpha)^{-g}$ with the microscopic lengthscale
$\alpha=v_F/2g\omega_c$ does not depend on impurity properties.

The effect of the external voltage can now
be read off from Eqs.~(\ref{gen}) and (\ref{dissac}). 
First, the average density [the first two terms in Eq.~(\ref{bosondens})]
is in general discontinuous at the impurity location
due to particle reflection,
\begin{equation} \label{backgr}
\bar{\rho} = \frac{k_F}{\pi} - 
\frac{e\varphi}{2\pi v_F}\, {\rm sgn}(x)\;.
\end{equation}
For the noninteracting case, $g=1$, one can show 
from the exact solution of the equivalent Schr\"odinger equation
that $\varphi$ is the usual {\em four-terminal voltage} measured near the 
barrier\cite{butt}. In that case,
$\varphi= R U$, where $R=1/(1+\lambda^{-2})$
is the reflection coefficient of the barrier.  

Second, in the absence of a scatterer, $\lambda=0$, one always 
finds $\varphi=0$ (see below),
and the right- and left-moving densities are spatially homogeneous
along the wire. Since they are determined by Eq.~(\ref{bc1}), the 
current is 
\[
I= \frac{i}{\sqrt{\pi}}\left\langle
 \frac{\partial \theta}{\partial \tau} \right\rangle=
e v_F(\langle\rho_+\rangle - \langle\rho_-\rangle) = \frac{e^2}{h} \,U\;.
\]
This yields the perfect two-terminal conductance $e^2/h$ 
in agreement with recent theoretical work \cite{pono}
and an experimental study of a quantum wire \cite{jap}.
The conductance $g e^2/h$ discussed in Ref.\cite{kane92}
is not the two-terminal conductance but a low-frequency 
microwave conductance.

To describe coupling to an external voltage, previous studies have 
often added a term to the Hamiltonian of the form\cite{kane92,weiss}
\begin{equation} \label{old}
\tilde{H}= e \tilde{\varphi}\, \theta(0) / \sqrt{\pi}
= \frac{e\tilde{\varphi}}{2\pi} \, q\;,
\end{equation}
where $\tilde\varphi$ is the ``voltage drop''\cite{foot1}.
From Eq.~(\ref{dissac}), if one tentatively 
identifies $\varphi$ with $\tilde\varphi$,
one observes that Eq.~(\ref{old})
should be modified by a factor $1/g^2$. This factor can be understood
in terms of the
interaction energy of density fluctuations with the nonequilibrium
background charge (\ref{backgr}) deviating from $k_F/\pi$,
\begin{eqnarray*}
&& \int dx \int dx' \,
\frac{1}{\sqrt{\pi}} \partial_x \theta(x) \, U_c(x-x')
 \left( \frac{-e\varphi\,{\rm sgn}(x')}{2\pi v_F} \right) \\
&& = -\frac{e\varphi}{2\sqrt{\pi}}\left(\frac{1}{g^2}-1\right)
 \int dx \, {\rm sgn}(x) \partial_x \theta(x) \;,
\end{eqnarray*}
where $U_c(x-x')$ is the screened Coulomb interaction, and the
last line holds for a Luttinger liquid.
Furthermore, it should be noted that in Eq.~(\ref{dissac}) the cosine part 
due to the impurity  has acquired a term linear in 
time, which is reminiscent of the Josephson relations.
This shows that the external voltage cannot be fully incorporated
by simply adding a term like Eq.~(\ref{old}) to the Hamiltonian.
In general, it is necessary to treat external reservoirs via 
boundary conditions \cite{footn}.  

From the above considerations, we see that the 
action (\ref{dissac}) describes
a voltage-biased 1D quantum wire containing a scatterer
for the entire range of parameters.
When evaluating Eq.~(\ref{gen}), we still have to average
over the zero mode of the Lagrange multiplier field.
Hence $\varphi$ is generally a {\em fluctuating} quantity.
In the two limiting cases of transmission zero and one,
the fluctuations do vanish. In the latter case, $\lambda=0$, 
 the $q$ average is Gaussian, and one finds $\varphi=0$ due to
the infrared singularity of the first term in the action\cite{kane92}.
On the other hand, for $\lambda\to \infty$, the
cosine term in Eq.~(\ref{dissac}) strictly enforces $\varphi=U$.
In the following, we will only consider the two fixed-point values
$\varphi=0$ and $U$ corresponding to a very small and a very high barrier, 
respectively. Near these limiting cases, further analytic progress
can be made.

We  start by calculating the nonequilibrium electron density. Equivalent to
an explicit real-time calculation, we first analytically continue to 
imaginary values of $U$ and $\varphi$, and after performing the $q$
average, we rotate back to real values of $U$ and $\varphi$.
Let us first study the case of a very {\em weak scatterer}, such that we can
put $\varphi=0$ and then use perturbation theory in the impurity
strength. The antisymmetric charge distribution 
\[
q(x)= -e\, [\langle\rho(x)\rangle-\langle\rho(-x)\rangle]/2\qquad 
(x>0)
\]
can be computed by expressing the density operator
(\ref{bosondens}) in terms of the generating functional (\ref{gen}).
Lowest-order perturbation theory in $\lambda$ yields
\begin{eqnarray} \label{pert2} 
q(x) &=& \frac{e\lambda k_F}{\pi} \,\sin(2k_F x) 
\frac{\sqrt{\pi}}{2\Gamma(g)}
\, (x/\alpha)^{-(g-1/2)} \\ \nonumber
&\times& (eU/\omega_c)^{g-1/2} \, J_{g-1/2} (geUx/v_F) \;,
\end{eqnarray} 
where $J_\nu(x)$ is a Bessel function of the first kind\cite{abram}.
Near the impurity, for $x\ll (eU/v_F)^{-1}$, 
from properties of the Bessel function,
\[
q(x) = \frac{e\lambda k_F}{2\sqrt{\pi}\Gamma(g) \Gamma(g+1/2)} 
\,\sin(2k_F x) \, (eU/2\omega_c)^{2g-1} \;.
\]
The asymmetric charge mode $q(x)$ is $2k_F$-periodic but decays
only on the lengthscale $v_F/eU$. There is no localized charge sitting on
the impurity. However, the $x$-integration over (\ref{pert2})
gives a finite total charge, and hence a finite nonlinear capacitance
$C=Q/U$,
\begin{equation} \label{caplutt}
C(U) = \frac{e^2 \lambda/\omega_c}{8\sqrt{\pi}\, \Gamma(g) \Gamma(g+1/2) }
\left(\frac{eU}{2\omega_c}\right)^{2g-2}  \;.
\end{equation}
This lowest-order perturbational result in the weak-scattering regime
breaks down for small voltages, $eU/2\omega_c \ll \lambda^{1/(2-2g)}$.

In the opposite case of a {\em strong scatterer}, $\lambda\gg 1$, we can
put $\varphi=U$. From Eq.~(\ref{gen}), the $2k_F$-part of
the asymmetric charge mode takes the form
\[
q_{2k_F}(x)= - \frac{e k_F}{\pi} \cos(2k_F x) \sin(eUx/v_F) (x/\alpha)^{-g}\;,
\]
which implies a finite total charge $Q_{2k_F}$. This charge turns out to be 
linear in $U$, and therefore one has a finite $U=0$ capacitance
\begin{equation} \label{cap22}
C_{2k_F}=
 \frac{e^2}{4\pi\omega_c} g^{-g} \Gamma (2-g) \sin\left[ \frac{\pi}{2}(1-g)
\right] \;.
\end{equation}
However, $q_{2k_F}(x)$ does not include the constant term 
$\bar q= eU/2\pi v_F$
coming from the change in background density (\ref{backgr}). 
That term leads to a charge
$Q_0= L e^2 U/4\pi v_F$, which diverges with the system length $L$.
The applied voltage polarizes the 
capacitance  between the wire and the metallic
screening gate. This large shunt capacitance renders  the observation of
charging effects, i.e.\ of $C_{2k_F}$, impossible
for a single scatterer. A similar situation is 
encountered for a single tunnel junction coupled to metallic leads, 
where charging effects are normally absent\cite{sct}. 

For an island formed by {\em two} strong impurities,
the $2k_F$ capacitance is observable since 
the capacitance of the island is not affected by the shunt
capacitance between the remaining wire and the gate. Taking two
impurities at $x=\pm R/2$ and applying our
boundary condition (\ref{bc}), we now have two zero modes from the
respective Lagrange multipliers. One ($\varphi$) corresponds to the 
four-terminal voltage found in the single-impurity case,
and the other ($\varphi^{}_G$) corresponds
to a gate voltage applied to the island $-R/2 < x < R/2$. 
While again analytic results are not available for the entire
range of parameters, it is possible to calculate the total charge $Q_I$
sitting on the island in the limit of large barriers 
and for $k_F R \gg 1$. We obtain 
\begin{eqnarray} \nonumber
Q_I &=& \frac{eR}{\pi}\, (k_F+e\varphi^{}_G/2 v_F) 
+ 2\, C_{2k_F} \varphi^{}_G \\ &-& 
e\, \Gamma(1-g) \sin[\pi
(1-g)/2]/\pi g^g 
 \;, 
\end{eqnarray}
where $C_{2k_F}$ is given in Eq.~(\ref{cap22}). 
The first term $\sim R$ arises due to the slow component of the 
density operator (\ref{bosondens}), while the remaining two terms
 come from the $2k_F$ component.
The resonant tunneling condition can then be derived
by noting that $Q_I/e$ is confined to integer values in the
large-barrier limit under consideration here. Hence, the spacing of the
resonances as a function of $\varphi^{}_G$ is found to be
\begin{equation}
e \Delta \varphi^{}_G = ( R/2\pi v_F + 2 C_{2k_F}/e^2 )^{-1} \;.
\end{equation}
Therefore $C_{2k_F}$ leads to an experimentally measurable
resonance shift compared to previous
results \cite{kane92} which neglected charging effects in
interacting 1D metals.

Let us finally comment on the issue of {\em electroneutrality}
in a Luttinger liquid.
The spatial change in the background charge density (\ref{backgr})
induces an influence charge density of opposite sign on the 
metallic gate, such that overall electroneutrality is maintained.
However, the Luttinger liquid interaction is not able to 
enforce electroneutrality within the 1D quantum wire alone.
In the complete absence of a gate, since
there is no internal screening within the wire,
the Coulomb potential becomes long-ranged,
$U_c(x-x')\sim |x-x'|^{-1}$. From the 
Euler-Lagrange equation corresponding to Eq.~(\ref{parti}) and the boundary 
condition (\ref{bc}), one finds immediately that then $\varphi=0$. 
As a consequence, for a long-ranged Coulomb potential, electroneutrality
is maintained automatically within the wire. 

In conclusion, the theoretical description of a voltage-biased
1D quantum wire (Luttinger liquid) containing elastic potential scatterers
has been given.
Our boundary condition method can easily be adopted to the case of
 more than one channel (e.g., the spin-$\frac12$ case), and it 
can straightforwardly be generalized 
to a real-time and finite-temperature formalism.


\begin{references}
\bibitem{foot1} The expression (\ref{old}) follows from 
$\tilde{H}= \int dx \, \mu(x) \rho(x) =
\int dx (-e\tilde{\varphi}\, {\rm sgn}(x)/2) \, \partial_x 
\theta(x)/\sqrt{\pi}$, which tacitly assumes that the electro-chemical
potential $\mu(x)$ is independent of $\rho(x)$.
\end{references}
\end{document}